\documentclass[twocolumn,prb,floatfix,superscriptaddress]{revtex4}

\usepackage{amssymb}
\usepackage{bm}
\usepackage[usenames, dvipsnames]{color}
\usepackage{mathrsfs}
\usepackage{graphicx}
\usepackage{amsmath}
\usepackage{enumitem}
\usepackage{natbib}
\usepackage{float}
\usepackage{verbatim}
\usepackage{color}

\newcommand{\imx}{\text{Im}\chi}

\newcommand{\tix}{\tilde{\chi}}

\newcommand{\beq}{\begin{equation}}
\newcommand{\eeq}{\end{equation}}
\newcommand{\beqr}{\begin{eqnarray}}
\newcommand{\eeqr}{\end{eqnarray}}
\usepackage{color}

\begin{document}

\title{Noise-induced Renyi entropy flow of a quantum heat engine}
\author{Mohammad H. Ansari}
\affiliation{Kavli Institute of Nanoscience, Delft University of Technology, P.O. Box 5046, 2600 GA Delft, The Netherlands}

\date{\today}
\begin{abstract}
Entropy is one of the central quantities in thermodynamics, whose flow between two systems determines the statistics of energy transfers. In quantum systems entropy is non-linear in density matrix whose time evolution is cumbersome.  Using recent developments in the Keldysh formalism for the evolution of nonlinear quantum information measures (Phys. Rev. B 91, 174307 (2015)),  we study the flow of von Neumann and Renyi entropies in a generic four-level quantum system that is weakly coupled to  equilibrium heat engines. We show that noise-induced coherence has significant influence on the entropy flow of the quantum heat engine.  We determine analytical optimization of couplings for the purpose of  designing optimal artificial energy transfer systems. \end{abstract}

\maketitle


\section{Introduction}
\label{intro}

Thermodynamics  governs a set of laws to relate different types of energy transfers in large systems.  As the system size decreases the energy fluctuations  become more relevant; thus  the  laws cannot be taken for granted in small scales.\cite{Maruyama} In small scales the laws have been reformulated using probabilistic stochastic phenomena such as those taking place in biomolecular processes.\cite{{Dorfmana},{sambur},{cho},{schaller}} However not all phenomena in small scale can be described by probability theory. The quantization of thermodynamics is essential even in scales much larger than atomic scale\cite{Horodecki}. In the past decade modern condensed matter successfully achieved to engineer controllable quantum phenomena, such as macroscopic quantum dissipation\cite{{vidal},{dur}}. This has been the driving force behind recent progress in mesoscopic physics to build novel quantum devices such as quantum bits (qubits) for  computation\cite{{clarke},{wendin}}. 

More recently experimental mesoscopic physics has entered the realm of quantum thermodynamics by changing demands from quantum computational devices \cite{{jarzynski},{koski},{seifert},{toyabe},{esposito},{pekola}}. However on theoretical basis there are not much known about how to consistently quantize thermodynamical quantities.  One of the central quantities is entropy, which is nonlinear in density matrix $\hat{\rho}$, i.e. $S=-\textup{Tr}\hat{\rho}\ln \hat{\rho}$. The quantization of such a nonlinear quantum information measure is not straightforward\cite{nazarov11};  therefore its derivatives are not necessarily related to the expectation value of any physical operator\cite{Amico}. In open quantum system, the usual approach to evaluate entropy is to take the following kinematical steps:  1)  reduced density matrix $\hat{\rho}_r(t)$ at arbitrary time $t$ is calculated from non-unitary quantum evolution equation \cite{{nazarovbook},{Keldysh},{Kamenev}}, and 2) the solution is substituted in the definition of entropy to obtain $S$ at time $t$ .  However, recently it has been understood that this $S(t)$ excludes most of energy relaxations between the open quantum system and its environment\cite{{Ando},{Campisi},{ansari1}}. To understand the inconsistency, let us reformulate von Neumann entropy in the equivalent form of Renyi entropy;  $S=-\lim_{M\to 1} dS_M/dM$ with $S_M= \textup{Tr}_r \{\rho_r\}^M$ being the Renyi entropy of positive degree $M$.\cite{renyi}  We developed a formalism based on Keldysh technique to consistently evaluate the dynamical quantization of $\{\hat{\rho_r}\}^M(t)$.\cite{{ansari1},{ansari3}}   Therefore entropy in quantum physics can be calculated by taking the following steps: 1) evaluating $\{\hat{\rho}_r\}^M(t)$ from the extended Keldysh formalism in multiple parallel worlds\cite{ansari1},   2) substituting the solution in the definition of Renyi entropy, and finally 3) taking the analytical continuation of the Renyi entropy  in the limit of $M\to 1$. Our formalism is  applicable to consistently calculate nonlinear quantum information measures  and also evaluate multiple qubit decoherence in strong coupling limit.\cite{{ansari3},{ansari4}}

Nonlinear quantum information measures, such as entropy, are unphysical quantities as they  cannot be determined from immediate measurements; instead their quantification is equivalent to determining the density matrix. This requires reinitialization of the density matrix between many successive measurements. Direct measurements of density matrix for a probe environment requires characterization of reduced density matrix of an infinite system, which is a rather nontrivial procedure and needs the complete and precise reinitialization of the initial density matrix. Recently in Ref. [\onlinecite{ansari2}] we proved a correspondence in weakly coupled generic open quantum systems between entropy flow and the flow of physical quantities, namely the full counting statistics of energy transfers\cite{Kindermann}. Expectedly the quantum features of entropy should be measurable in quantum heat engines (QHE). These engines are made of several heat bath kept at different temperatures and connected via a few degrees of freedom. They convert incoherent photon energy of thermal baths to coherent emission \cite{schully}. Examples are light-harvesting systems\cite{{cao},{whaley}}, quantum photocells\cite{schullypc}, photosynthetic organic cells \cite{Dorfmana}, and laser heat engines.\cite{{sambur},{cho},{schaller}} Novel quantum phenomena have been  captured in the engines such as lasing without inversion \cite{schullylasing}, extraction of work from a single thermal reservoir \cite{schullysingle}, and enhancement of the output power\cite{schullypc}.  

\begin{figure}
\begin{center}
\includegraphics[scale=0.4]{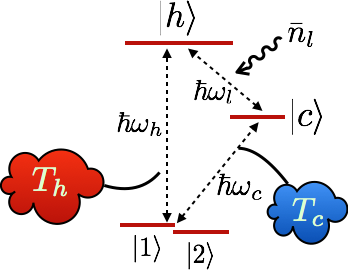}
\caption{(Color online) A quantum heat engine.  The degenerate levels $|1,2\rangle$ are resonantly coupled to two excited levels $|h\rangle$ and $|c\rangle$  by two thermally populated field modes with hot ($T_h$) and cold ($T_c$) temperatures. Levels $|h,c\rangle$  are coupled by a non-thermal cavity mode. }
\label{fig1}
\end{center}
\end{figure}

Previously we studied the flow of entropy in some QHEs and showed new features.\cite{{ansari1},{ansari3},{ansari4},{ansari2}} Here we consider a practical QHE in which heat dissipation take place via four energy levels two of which are degenerate ground states, see Fig. (\ref{fig1}).  Such an engine has been recently modelled by Scully et. al. in Refs. [\onlinecite{{schully},{schullypc},{schullylasing},{schullysingle}}] and showed to be able to increase the engine output power due to the contribution noise-induced quantum coherence\cite{schullypc}, a pure quantum effect. Examples of such engines are in photovoltaic cells consisting of quantum dots sandwiched between doped semiconductors\cite{{khoshnegar},{kirk}} and lasing heat engines\cite{nikonov}. Increasing the interaction couplings makes the energy fluctuation to become deviated from Poissonian distribution\cite{Rahav}.  Here we restrict our study to weak coupling regime where the full dynamical quantization of non-linear measures have been understood \cite{ansari3}. We compute the flow of  entropy in the QHE and show interesting quantum features in the entropy flow.  Taking into account of the behaviour of entropy flow in the QHE, we are able to optimize coupling and temperature in designing optimal artificial energy transfer systems.

The paper is organized as follows: in Section (\ref{model}) we present a model  for quantum system that exchanges photon with some heat baths. Using the extended Keldysh formalism we take a diagrammatic approach to directly calculate the Renyi entropy flow for  exchanged photons.  In section (\ref{qhe}) we calculate the flow of Renyi entropy for a 4 level quantum heat engine and discuss some of its interesting physics. Results are summarized in section (\ref{discussion}).

\section{The Model and Method}
\label{model}

Our aim in this section is to take  pedagogical steps to evaluate the renyi entropy flow in the quantum heat engine that contains the simplest heat bath using the extended Keldysh formalism in multiple parallel worlds\cite{ansari4}.  Although the abstract form of the entropy flow in a generic system can be found elsewhere\cite{{ansari1},{ansari3},{ansari4}}, however the explicit form of new correlators in terms of couplings will provide an opportunity to expect novel quantum phenomena emerge from different set of couplings. This will also motive a curious reader to study other types of baths and electron reservoirs.  

Let us label a set of discrete energy levels by $|x\rangle$ with the corresponding energy $E_x$. The energy levels and photons in the heat bath $\alpha$ are weakly coupled, this enables to calculate the entropy from perturbation theory.   The Hamiltonian is $H=H_{0}+H_{int}$ with non-interacting part $H_0$ as the sum of system and photon reservoirs,  
\beqr \nonumber
\hat{H}_{sys}=\sum_{x} E_x|x\rangle \langle x|, && \\ 
\hat{H}_{\alpha}= \sum_{{\bm{q}},\alpha}\hbar \omega_{{\bm q},\alpha} \hat{b}_{{\bm q},\alpha}^\dagger \hat{b}_{{\bm q},\alpha}, &&
\label{eq.H0}
\eeqr
 with $\hat{b}_{{\bm q},\alpha}$ ($\hat{b}_{{\bm q},\alpha}^\dagger$) being annihilation (creation) photon operator with  momentum ${\bm q}$ in the reservoir labelled $\alpha$. The interaction Hamiltonian is 
\beqr
\label{eq. V} \nonumber
&& \hat{V}= \sum_\alpha \sum_{xx'} \left|x\rangle\langle x'\right|X^{(\alpha)}_{xx'}\left(t\right) ,\\
&& \hat{X}^{(\alpha)}_{xx'}(t)=\hbar \sum_{{\bm{q}}}c_{xx',{\bm{q} \alpha }}\hat{b}_{{\bm{q} \alpha}}\exp(-i\omega_{{\bm{q} \alpha}} t)+h.c. 
\eeqr
with the complex coupling energy $c_{xy,q}$.

We assume adiabatic switching of the perturbation such that far in the past $t\to -\infty$ the coupling is absent, and the density matrix is the direct product of subsystems. The coupling slowly grows achieving actual values at a long time in the past of present time $t$. The time evolution of density matrix $\rho_{xy}$ in the quantum system formally takes place as 
\beq
\hat{\rho}(t)=Te^{i\int_{\infty }^t d\tau \hat{V}(\tau)}\hat{\rho}(-\infty) \bar{T}e^{i\int_{-\infty}^t d\tau \hat{V}(\tau)},
\label{rhot}
\eeq
with $T$ ($\bar{T}$) being (anti-) time ordering operator.  Using standard Keldysh formalism one can expand the non-unitary operators in terms of $\hat{X}$ operators in all orders; this sets the time ordering along the Keldysh contour that assumes opposite timing for bra and ket state evolutions. From the perturbative expansion in the second order of density matrix, one can see the density matrix of quantum system can be evaluated from the following quantities $\int_{0}^{\infty}S_{xy,zt}(\tau)  \exp(i\omega\tau) d\tau $ with $S_{xy,zt}(\tau) = \langle \hat{X}_{xy}\left(t\right)  \hat{X}_{zt}\left(t+\tau\right)\rangle $ being a photon exchange correlator and $\langle \cdots \rangle$ denotes trace over photon states re-summed over all such states, see Appendix \ref{app1} for details.    The integral for thermally equilibrium baths can be simplified to $(1/2) S_{xy,zt}\left(\omega\right)+i\Pi_{xy,zt}\left(\omega\right)$ with the Fourier transform of the correlator $S_{xy,zt}$ and $\Pi_{xy,zt}$ defined  as follows: 
\beqr \nonumber
S_{xy,zt}\left(\omega\right) & \equiv & 2\pi \hbar^2\sum_{q}c_{xy,q}c_{tz,q}^{*}\langle b_{q}b_{q}^{\dagger}\rangle\delta\left(\omega+\omega_{q}\right)\\  &&\quad \ \   +c_{yx,q}^{*}c_{zt,q}\langle b_{q}^{\dagger}b_{q}\rangle\delta\left(\omega-\omega_{q}\right) \label{eq. P}\\  \nonumber
\Pi_{xy,zt}\left(\omega\right) & \equiv & \hbar^2 \sum_{q}c_{xy,q}c_{tz,q}^{*}\langle b_{q}b_{q}^{\dagger}\rangle\left(\frac{1}{\omega+\omega_{q}}\right)\\ && \quad +c_{yx,q}^{*}c_{zt,q}\langle b_{q}^{\dagger}b_{q}\rangle\left(\frac{1}{\omega-\omega_{q}}\right) 
\label{eq. Q}
\end{eqnarray}
Depending on the order of the $\hat{X}$ operators there are other possible correlators. The correlators in a general non-equilibrium system are independent, however the state of thermal equilibrium adds a strong constraint such that all correlators can be evaluated from dynamical susceptibilities using  the so-called Kubo-Martin-Schwinger (KMS) relations [22].  Given the definition of dynamical susceptibility in the environment as $\chi_{xy,zt}\left(\omega\right)  =  (-i/\hbar) \int_{-\infty}^{0} \langle [\hat{X}_{xy}\left(t+\tau\right),\hat{X}_{zt}\left(t\right) ]\rangle \exp({-i\omega\tau})d\tau$, KMS relation allows to determine $S_{xy,zt}$ from the following part of the dynamical susceptibility $\tilde{\chi}_{xy,zt}= (\chi_{xy,zt}-\chi^*_{tz,yx})/i$; i.e. $S_{xy,zt}(\omega)= \hbar \bar{n}(\omega/T) \tilde{\chi}_{xy,zt}(\omega)  $ with $\bar{n}(E/T)=(\exp(E/k_BT)-1)^{-1}$ being the Bose distribution function. 

\emph{Renyi entropy}: Evaluating R\'{e}nyi entropies requires time evolution of integer powers of density matrix.   Consider a closed system with total---namely  \emph{world}---density matrix $\rho$ made of two interacting subunits $A$ and $B$. The reduced density matrix for system $A$ is  $\rho_A=\textup{Tr}_{B} \rho$. The Renyi entropy of system $A$ is  $\ln S_M^{(A)}=\ln \textup{Tr}_A \{ \rho_A ^M\}$. If the two systems do not interact, the entropies are conserved $d\ln S_M^{(A,B)}/dt=0$; however for interacting heat baths in thermal equilibria, a steady flow of entropy is expected from one heat bath to another one. This is similar to the steady flow of charge in an electronic junction that connects two leads kept at different chemical potentials\cite{ansariqp}. Defining the Renyi entropy flow in $A$ as $F_M^{(A)}=-d\ln S_M^{(A)}/dt$,  there is a conservation law for  $F_M^{(A+B)}$; however due to the inherent non-linearity $F_M^{(A)}+F_M^{(B)}\neq 0$ and equality holds only approximately, subject to volume dependent terms.\cite{nazarov11}

\begin{figure}[htbp]
\begin{center}
(a)\includegraphics[scale=0.16]{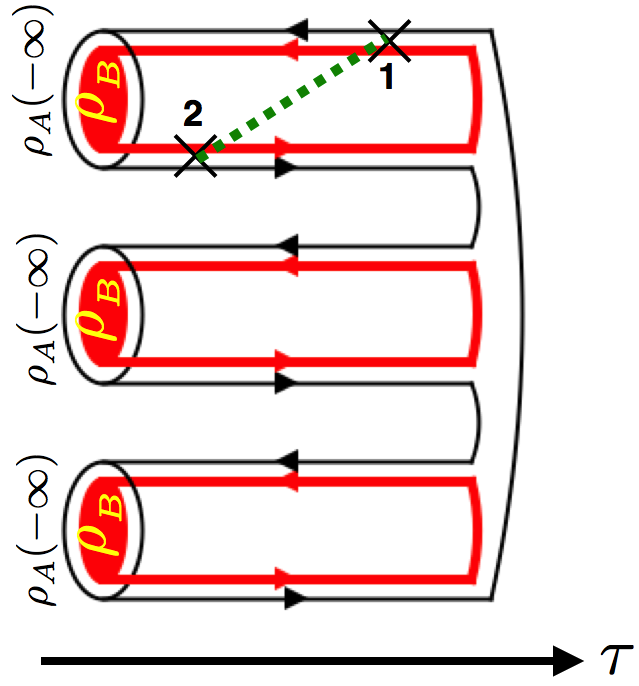}
(b) \includegraphics[scale=0.16]{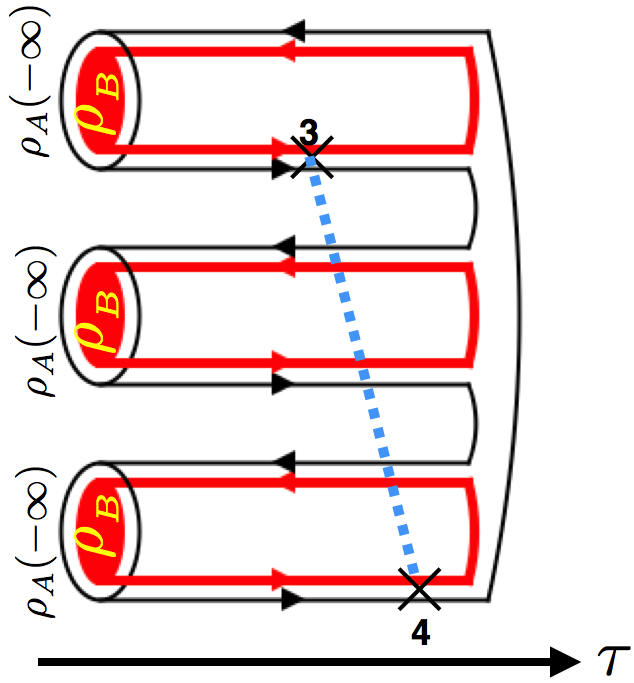} 
\caption{(Color online)  Two typical diagrams in the the calculation of Renyi entropy flow in system A for $M=3$. Time evolves from left to right The contours for system A (B) are depicted as thin black (thick red) lines. There are three worlds contributing and the expectedly the density matrix of each system B is traced over within each world. The interaction between A and B is denoted by a cross.  The diagram (a) shows two interactions at 1 and 2 taking place inside a single-world. This in the diagram (b) is labeled with 3 and 4 and take place across two different world. These latter types of interaction are non-trivial quantum contribution to the evolution of quantum information measures with non-linear dependence on density matrix. }
\label{diags}
\end{center}
\end{figure}

Computing the flow of Renyi entropy can be done by extending  standard Keldysh formalism from one unitary evolution of density matrix to multiple parallel  evolution for several density matrices, all in the time interval $(-\infty, t]$.\cite{ansari1} This can be diagrammatically drawn using $M$ parallel bra and ket-contours in Fig. (\ref{diags}).  In the second order perturbation terms of $d\rho^M/dt$ includes the second order term $M (\delta^{(2)}\rho ) \rho^{M-1}$. The novelty of this formalism is that it helps to see there are contributions from single order terms in the following form  $ (\delta^{(1)} \rho) \rho^n (\delta^{(1)}\rho) \rho^{M-n-2}$. Fig. (\ref{diags}) shows two typical diagrams of exchanging photons between A and B: in (a) the two interactions occur in a single world, and in (b) the two interactions take place across two different worlds. The diagram (a) corresponds to the following term in the $M=3$ Renyi entropy flow:  $ \textup{Tr}\{ V_1(t+\tau)\rho_A V_2(t)) \rho_A^{M-1}\}/dt $, and (b) is equivalent to $ \textup{Tr} \{\rho_A V_3(t) \rho_A^2 V_4(t+\tau) \}$. The former indicates that once a photon is exchanged within a world the outcome state  kinematically passes through other $M-1$ worlds. The latter allows for the exchange of photon between two different worlds.  

Let us consider $A$ being a thermal heat bath at temperature $T=1/k_B\beta$ with diagonal density matrix on the eigenmodes $|\{n_{i}\}\rangle\equiv|n_{1},\cdots,n_{q},\cdots\rangle$
with probability $\mathcal{P}\left(\left\{ n_{i}\right\} \right)=\prod_{i}p\left(n_{i}\omega_{i}\right)$
with $p\left(n_{i}\omega_{i}\right)=\exp\left(-\beta\hbar\omega_{i}n_{i}\right)/Z\left(\omega_{i}\right)$ and $Z(\omega_i)$ being the partition function of mode $\omega_i$. Consider the creation  operator of a photon $\hat{b}\dagger$ to take place in a world at a moment. This changes the state of reservoir as long as the photon is not annihilated. The annihilation $\hat{b}$ occurs after the photon kinematically passes through $n$ copies of density matrix and after which the states in both A and B in the energy representation turn back to the initial state. The generalized correlators after some simple algebra becomes
\beqr 
\frac{ \left\langle \hat{b}_{q}\rho_{\textrm{env}}^{n} \hat{b}_{q}^{\dagger}\rho_{\textrm{env}}^{M-n} \right\rangle }{ \left\langle \rho_{\textrm{env}}^{M} \right\rangle }   & = & \frac{\exp({\hbar\omega_{q}(M-n)/k_BT})}{\exp({\hbar\omega_{q}M}/k_BT)-1}
\label{eq: corr b rho bdag rho}
\eeqr

The  correlator can be substituted in the extended time evolution of $d \textup{Tr}\{(\rho_A)^M\}/dt$ whose evaluation requires the  following generalized correlator:
\beq
S^{n,M}_{xy,zt}(\tau)=\textup{Tr}_b\{\hat{X}^{(b)}_{xy}(t)\rho_b^n \hat{X}^{(b)}_{zt}(t+\tau)\rho_b^{M-n} \}/\textup{Tr}_b\{\rho_b^M\}
\label{eq.corr}
\eeq

Using the definition of $\hat{X}$ in Eq. (\ref{eq. V}) and simplifying the integral will result in the following explicit form of photon general correlator 
\beqr \nonumber
&&S^{n,M}_{xy,zt}\left(\omega\right)  =  2\pi\hbar^2 \sum_{q}\left\{ c_{xy,q}c_{zt,q}^{*}\frac{e^{\beta\hbar\omega_{q}\left(M-n\right)}}{e^{\beta\hbar\omega_{q}M}-1}\delta\left(\omega+\omega_{q}\right)\right. \\  && \qquad  \ \  \qquad \qquad \left. +c_{xy,q}^{*}c_{zt,q}\frac{e^{\beta\hbar\omega_{q}n}}{e^{\beta\hbar\omega_{q}M}-1}\delta\left(\omega-\omega_{q}\right)\right\} 
\label{eq: S(w,N)}
\eeqr
that satisfies the important relation $S^{n,M}_{xy,zt}\left(-\omega\right)  =  S^{M-n,M}_{zt,xy}\left(\omega\right) $. 

 In the thermal equilibrium one can obtain generalized KMS relation between the generalized correlators and dynamical susceptibility $\tilde{\chi}$ defined above, 
\beq
 S_{xy,zt}^{n,M}\left(\omega\right)= \hbar  \bar{n}(M\omega /T)   \exp({\hbar n \omega }/k_BT)  \tilde{\chi}_{xy,zt}\left(\omega\right)
\label{eq. KMS}
\eeq

Similar to the above discussion after Eq. (\ref{rhot})  the time evolution of the operator $\rho(t)^M$ in the time interval $(-\infty,t]$ can be determined through the evaluation of integrals $\int_0^\infty d\tau \left\langle X_{xy}\left(t\right)\rho_b^{n} X_{zt}\left(t+\tau\right) \rho_b^{M-n} \right\rangle e^{i\omega\tau}/{\left\langle \rho_{\textrm{env}}^{M} \right\rangle }$. Simple algebra shows that the integral is $(1/2) S_{xy,zt}^{n,M}(\omega) + i \Pi_{xy,zt}^{n,M}(\omega)$, with $ \Pi_{xy,zt}^{n,M}(\omega)= (1/2\pi) \int dz S_{xy,zt}^{n,M}(z)/(\omega-z)$

We implement the extended Keldysh formalism for the analysis of R-flow. Detailed analysis with all diagrams that can be seen in  Appendix B of [\onlinecite{ansari1}]  and using the generalized KMS relation (\ref{eq. KMS}) will result the following flow of Renyi entropy:
\beqr \nonumber
 F_M   &=& \frac{ M}{\hbar}\sum_{yy'} \frac{\bar{n} \left(M\omega_{yy'}\right)}{\bar{n}\big(\left(M-1\right)\omega_{yy'}\big)}  \left\{ Q_{yy'}^{(i)} - Q^{(c)}_{yy'}\right\} 
   \label{eq. rflow gen}
\eeqr
with
\beqr \nonumber
Q^{(i)}_{yy'}   &=&   \sum_{x'}\rho_{x'y}  \tilde{\chi}_{y'x',yy'}\left(\omega_{yy'}\right)   e^{\beta \hbar \omega_{yy'}} \\ 
Q^{(i)}_{yy'}   &=&    \sum_{xx'}\rho_{x'x}\rho_{y'y}  \tilde{\chi}_{xx',yy'}\left(\omega_{yy'}\right)  \left( e^{\beta\hbar \omega_{yy'}} -1\right) 
  \label{eq. rflow}
\eeqr
where the  $\Pi$ part of the flow identically vanish in the summation. In Eq. (\ref{eq. rflow}) there are two types of flows contributing: (i) the incoherent flow  $Q^{(i)}$, for quantum leaps on energy levels, and (ii) the coherent flow $Q^{(c)}$ for the exchange of energy through the quantum coherence. We study a master equation approach in appendix (\ref{app0}).

Previously we have studied the R-flow  in some examples of quantum heat engines, such as engines made of a two level system in [\onlinecite{ansari1}] and a harmonic oscillator engine in [\onlinecite{ansari2}].  These systems exhibit non-trivial deviations from the classical thermodynamics. Below we apply the formulation to a quantum heat engine with four quantum states.

\section{A Four level QHE}
\label{qhe}

We consider the four-level quantum heat engine introduced by Scully \emph{et al.}\cite{schullypc}. The QHE consists of two nearly degenerate lower levels $|1\rangle$  and $|2\rangle$ with energy $E_1\approx E_2$, and two non-degenerate upper levels $|h\rangle$ and $|c\rangle$ with energies $E_h$ and $E_c$, see Fig. (\ref{fig1}). An example of such model is a laser heat engines in which noise-induced quantum coherence can help to increase the net emitted laser power.\cite{{schullypc},{schullylasing}} 

A heat bath at temperature $T_h$ ($T_c$) drives transitions between the ground states and $|h\rangle$ ($|c\rangle$) and the two excited levels are externally driven by a single-mode field with the Hamiltonian $\hat{H}_{sys-dr}=\Omega (\hat{b}_l^\dagger |c\rangle\langle h|+\hat{b}_l |h\rangle\langle c|)$,  $\hat{b}_l$ ($\hat{b}_l^\dagger$) being the annihilation (creation) operator for the cavity mode of frequency $\omega_{l}$ and $\langle \hat{b}^\dagger_l \hat{b}_l\rangle = \bar{n}_l $ and $\langle \hat{b}_l \hat{b}^\dagger_l \rangle = \tilde{n}_l\equiv \bar{n}_l+1$, $\bar{n}_l$ being the average number of photons in the cavity.

Detailed analysis of the four level reduced density matrix\cite{schullypc} show that the exchange of photon between ground states and higher levels generates steady quantum coherence $\rho_{12}$. Let us denote the non-vanishing elements of density matrix in a vector $\bm{R}=\{\rho_{11}, \rho_{22}, \rho_{hh}, \rho_{cc}, \textup{Re}(\rho_{12}) \}$, whose steady values  can be determined from the quantum evolution equation, see Appendix (\ref{app3}).  

We can compute  the flow of Renyi entropy in a probe heat bath, which we choose to be the hot bath at temperature $T_h$.   The flow of Renyi entropy for the QHE of four energy levels with nearly degenerate ground states from Eq.(\ref{eq. rflow gen}) is: 

\beqr \nonumber
 {F}^{(T)}_M&=& 
\frac{M \bar{n} \left[M\omega_h/T_h\right]  }{ \hbar \bar{n}\big[\left(M-1\right)\omega_h/T_h\big]} \bigg\{  \gamma  p_h  \exp({ E_h/k_BT_h})   \bigg. \\  &&  \bigg.  -\left(\tilde{\chi}_{h1} p_1+ \tilde{\chi}_{h2} p_2 + 2 \tilde{\chi}_{1h,h2} \textup{Re}\rho_{12}  \right)  \bigg\}   \label{eq. rflow ex}
\eeqr
with $\gamma \equiv \tilde{\chi}_{h1} + \tilde{\chi}_{h2} + 2 \tilde{\chi}_{1h,h2}$, $\tilde{\chi}_{h1,2h}=\sqrt{\tilde{\chi}_{h1} \tilde{\chi}_{h2}}$, and $p_x\equiv \rho_{xx}$.  One can see in Eq. (\ref{eq. rflow ex}) the quantum coherence influences the entropy.  Moreover the quantum coherence also influences state populations $\rho_{xx}$ for $x=1,2,c,h$---see Appendix \ref{app3}.

In Fig. (\ref{fig2}) the von Neumann entropy flow from/to the hot reservoir in the absence (presence) of quantum coherence, denoted by $dS'/dt$ ($dS/dt$), is plotted versus the temperature $T_c$ of the main environment for two different cases of $\tilde{\chi}_{c2}=0.1,1$ for a given coupling $\chi_{c1}=0.1$;  the parameters we use are from Ref. [\onlinecite{Rahav}].   Fig. (\ref{fig2}) inset shows that as $T_c$ increases the population of $|h\rangle$ increases, this increases the rate of photon absorption by the probe heat bath, thus a negative flow of entropy. In the lack of coherence, by setting $\tilde{\chi}_{1i,i2}=0$ for $i=h,c$, the onset temperature, where entropy flow vanishes, takes place at near a universal temperature $\sim 0.42$. Entropy flow is positive below the temperature and negative above it. This flow is similar to the heat current worked out in [\onlinecite{Rahav}].  However, in the presence of quantum coherence the universality of vanishing entropy flow is violated  as seen in $dS/dt$.  Another difference between $dS/dt$ and $dS'/dt$ in Fig. (\ref{fig2})  is that the presence of steady coherence the rate of entropy change not only shifts down but also falls much steeper, specially in the low $T_c$ limit. Interestingly quantum coherence slows down (speeds up) the rate of entropy change that flows out of (into) the probe environment.  

Our  formulation of entropy flow in  Eq. (\ref{eq. rflow ex}) shows that the sign change in the entropy flow takes place at ${ T_h^* }=E_h /\ln \left( (\sum_i \tilde{\chi}_{hi} p_i + 2 \tilde{\chi}_{1h,h2} \textup{Re}\rho_{12})  / \gamma  p_h \right)$, where the steady populations $p_i$'s and coherence $\textup{Re}\rho_{12}$ can be determined from the time evolution of the density matrix coupled to all heat baths, see Appendix \ref{app3}.   In the case of uniform couplings $\tilde{\chi}_{hi}=\tilde{\chi}_{h} $ for $i=1,2$ at high temperature limit $T_h\gg E_h/k_B$ the population of the ground states and the highest excited states becomes nearly similar. Expectedly the  von Neumann  entropy flows persistently out of the hot bath into the cold one and  monotonically decreases.

One can see in Fig. (\ref{fig2}) that  for the coupling $r_c\equiv \tilde{\chi}_{c2}/\tilde{\chi}_{c1}=10$  (the curve labeled $\tilde{\chi}_{c2}=1$) the quantum entropy increases below $T_c\sim 0.17$, as entropy flow is positive, and decreases above it. The case of  $r_c=1$ (the curve labelled $\tilde{\chi}_{c2}=0.1$) the entropy  monotonically decreases as it is negative at all temperatures; this is a persistent refrigerator. In the absence of quantum coherence a persistent  refrigerator is not possible.  From Eq. (\ref{eq. rflow ex})  condition for the negativity of entropy flow is: $(1+\sqrt{r_h})^2  \exp(E_h/k_BT_h) - \left(p_1+ r_h p_2 + 2 \sqrt{r_h} \textup{Re}\rho_{12}  \right)/p_h<0$ with    $r_h\equiv \tilde{\chi}_{h2}/\tilde{\chi}_{h1}$ being the ratio of degenerate state coupling to the highly excited level.  

From the details of equation of time evolution for $p_1$ and $p_2$ in Appendix \ref{app3} in the limit of $T_c \approx 0$, one can find the following steady population of states $p_i=p_h \exp(E_h/k_BT_h) + (\alpha_i/\bar{n}(E_h/T_h)) p_c - \lambda_i \textup{Re}{\rho}_{12}$ for $i=1,2$ and $\alpha_i=\tilde{\chi}_{ci}/\tilde{\chi}_{hi}$,  $\lambda_1=\sqrt{r_h}$ and $\lambda_2=1/\lambda_1$. Substituting these solutions into the above condition the negativity of entropy will become: ${p_h}/{p_c}<\eta  \left(1-\exp({-E_h}/{k_BT_h})\right) ({1+r_c})/{2\sqrt{r_h}}$. Substituting the steady solution for $p_c$ will lead to the following condition  for persistent refrigerator regime in the QHE: 
\beq
\label{eq.persis}
\frac{\sqrt{r_h}\eta}{1+r_c} + \frac{\chi_h}{\Omega^2 n_l}>{\frac{1}{2}(1-e^{\frac{-E_h}{k_BT_h}}})
\eeq
with $\eta\equiv  \tilde{\chi}_{h1}/\tilde{\chi}_{c1}$ and $\chi_h= \sqrt{\tilde{\chi}_{1h}\tilde{\chi}_{2h}}$. In the limit of uniform couplings the left side is minimally  $1/2$ which is in fact a maximal value for the right side. This just proved that for making a persistent cooling refrigerator non-uniform couplings are necessary between the degenerate states and any one of the heat baths. However note that making the couplings as different as possible does not guarantee to reach the persistent refrigerator condition. An example is the parameter used in the examples of Fig. (\ref{fig2}) one can see from Eq. (\ref{eq.persis}) that the persistent refrigerator for $r_h=2$ will be performed for the case $r_c=1$ but not for $r_c=10$. For low temperature probe environment $E_h\gg k_BT_h$, $\eta=1$, and large drive $\Omega^2n_l\gg \chi_h$ the inequality is simplified to $(1+r_c)^2<4r_h$.  

From Eq. (\ref{eq.persis}) one can see the persistent condition can be achieved by  reducing the Rabi frequency $\Omega$ and the number of cavity photons. Similarly the proper ratio of couplings does the same. For the uniform couplings $\tilde{\chi}_{hi}=\tilde{\chi}_{ci}\equiv \tilde{\chi}$ for $i=1,2$, in the high temperature limit $E_h\leq k_BT_h $ the condition is satisfied for $E_h/k_BT_h<1+2{\tilde{\chi}}/{\Omega^2 n_l} $. This indicates that decreasing the number of photons in the cavity will make the condition to be better satisfied.

\begin{figure}[htbp]
\begin{center}
\includegraphics[scale=0.12]{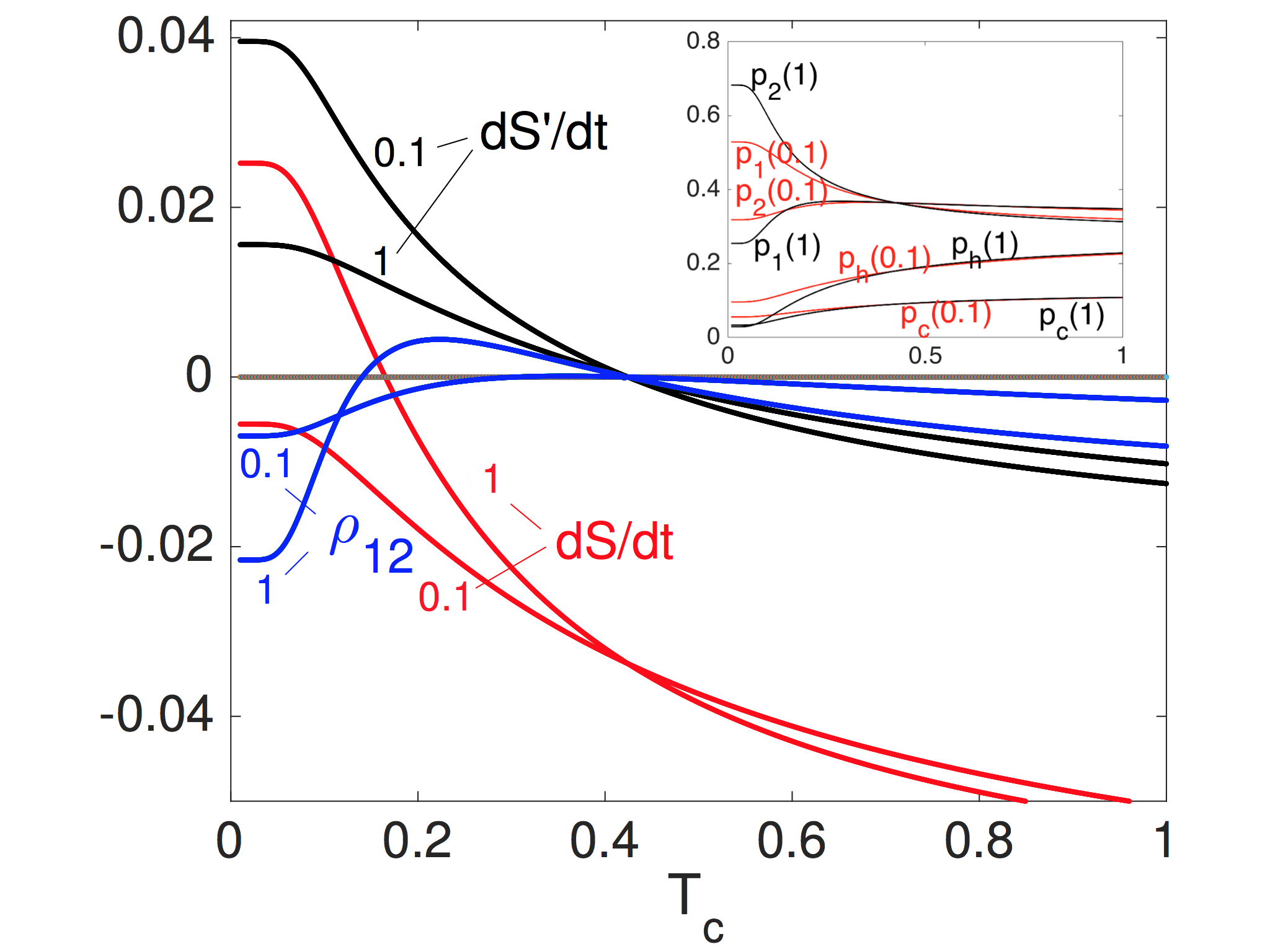}
\caption{(Color online)  The flow of  von Neumann  entropy for two values of $r_c=1,10$ in the presence of quantum coherence $dS/dt$---uppermost curves at right end of the box in black) and in the lack of coherence $dS'/dt$---the two middle curves at right end (in purple)---and the real part of quantum coherence $\textup{Re}\rho_{12}$---the lowermost two curves at right end of the box (in blue).  The fixed parameters are $\tilde{\chi}_{c1}=\tilde{\chi}_{h1}=\tilde{\chi}_{h2}/2=0.1$, $E_h=1.5$, $E_c=0.4$, $E_1=E_2=0.1$, $T_h=\Omega=n_l=\tau/5=1$. Inset: the probabilities $p_1, p_2, p_c, p_h$ for the two sets of parameters $\tilde{\chi}_{c2}=0.1$ (red) and 1 (black). }
\label{fig2}
\end{center}
\end{figure}

\section{Summary}
\label{discussion}

We calculated the Renyi entropy flow of bath-system entanglement in a four level quantum heat engine with doubly-degenerate ground states in the limit of weak interaction. To achieve a full dynamical quantization of entropy we use the extended Keldysh formalism in multiple parallel worlds.   Our exact calculation shows that quantum coherence directly influences the flow of  von Neumann  and Renyi entropies from a probe environment.  We found the onset temperature in the main environment  after which the quantum heat engine becomes a refrigerator that extract entropy from the main environment. The noise-induced coherence not only modifies the rate of entropy change, but also it allows certain conditions for persistent cooling even at ultracold temperatures.  Such non-trivial physics implies the importance of the concept of entropy flow to be further investigated.  

Let us briefly discuss how the physics of strong coupling regime can be studied.  Below I will describe two approaches for the development:  One can use the polaron transformation to incorporate the high-order system-bath interaction into the system dynamics. This transformation will change the generalized correlators of  heat baths as well as the Renyi entropies.  Alternatively, one can define the generalize density matrix $\mathcal{R}$ to include the density matrix of $M$ worlds, and extend the dynamical equation for  $\mathcal{R}(t)$. The solution is a set of eigensolutions proportional to $\mathcal{R}(t) \approx \exp (-\Gamma t )$. In strong coupling limit there is no stationary solution with zero $\Gamma$, instead the flow of Renyi entropy is $F_M=\Gamma_{0}$ with $\Gamma_{0}$ being the  closest eigenvalue to zero. 

\section{Acknowledgements}
The research leading to these results has received funding from the European Union Seventh Framework Programme (FP7/2007-2013) under grant agreement n° 308850 (INFERNOS).

\appendix

\section{Master equation approach}
\label{app0}

 Let us now  make an attempt to compute the Renyi entropy flow for the photon reservoir coupled to quantum system. The Master equation allows to characterize the time evolution of state populations $\mathcal{P}\left(\left\{ n_{i}\right\} \right)=\prod_i P({n_i})$. From definitions it is easy to find $F_M S_M =  -\sum_{i,n_i} d P({n_i})^M/dt$. Time evolution of $P(n_i)$  is determined from the semi-classical Master equation $dP(n_i)/dt= \sum_{m_i}-\Gamma_{m_i\to n_i} P(n_i)+\Gamma_{n_i \to m_i} P(m_i)$, with the transition rates $\Gamma_{m_i\to n_i}$ from a state with energy $m_i \omega_i$ to the state $n_i \omega_i$. Substituting this in the definition of the Renyi entropy flow one finds $F_M S_M =  \sum_{i,n_i} P({n_i})^{M-1}(\sum_{m_i}\Gamma_{m_i\to n_i} P(n_i)-\Gamma_{n_i \to m_i} P(m_i))$. From switching the labels $n_i$ and $m_i$ only in the second summation one gets the following general R-flow: 
\beq 
 F_M S_M = 
 \sum_{i,n_i,m_i} \Gamma_{m_i\to n_i}P(n_i) \left[P({m_i})^{M-1}-P({n_i})^{M-1}\right]
 \label{eq.master}
\eeq
For the choice of thermally equilibrium bath expectedly the result of eq. (\ref{eq.master}) in the weak coupling regime matches with the incoherent flow of Renyi entropy in Eq. (\ref{eq. rflow}).\cite{ansari1}

\section{Evolution equation}
\label{app1}

Using the interaction Hamiltonian defined in Eq. (\ref{eq. V}) we can evaluate the correlator integral using two functions $P$ and $\Pi$, 
\beq
\int_{0}^{\infty}\langle X_{xy}\left(t\right)  X_{zt}\left(t+\tau\right)\rangle e^{i\omega\tau}d\tau =  P_{xy,zt}\left(\omega\right)+iQ_{xy,zt}\left(\omega\right)
\label{eq. forward prop}
\eeq
with $\langle \cdots \rangle$ indicating trace over environment density matrix and 
\beqr \nonumber
P_{xy,zt}\left(\omega\right) & \equiv & \hbar^2\pi\sum_{q}c_{xy,q}c_{tz,q}^{*}\langle b_{q}b_{q}^{\dagger}\rangle\delta\left(\omega+\omega_{q}\right)\\  &&\quad \ \   +c_{yx,q}^{*}c_{zt,q}\langle b_{q}^{\dagger}b_{q}\rangle\delta\left(\omega-\omega_{q}\right) \label{eq. P}\\  \nonumber
\Pi_{xy,zt}\left(\omega\right) & \equiv & \hbar^2 \sum_{q}c_{xy,q}c_{tz,q}^{*}\langle b_{q}b_{q}^{\dagger}\rangle\left(\frac{1}{\omega+\omega_{q}}\right)\\ && \quad +c_{yx,q}^{*}c_{zt,q}\langle b_{q}^{\dagger}b_{q}\rangle\left(\frac{1}{\omega-\omega_{q}}\right) 
\label{eq. Q}
\end{eqnarray}

From the detail form of the functions $P$  in Eqs. (\ref{eq. Q}), one can show that for negative frequency the following relations hold
\begin{eqnarray}\nonumber
P_{xy,zt}\left(-\omega\right) & = &  e^{\beta\hbar\omega}\left[\pi\sum_{q}c_{xy,q}c_{tz,q}^{*}\frac{1}{e^{\beta\hbar\omega_{q}}-1}\delta\left(\omega-\omega_{q}\right)\right. \\ \nonumber && \left.\qquad \qquad +c_{yx,q}^{*}c_{zt,q}\frac{e^{\beta\hbar\omega_{q}}}{e^{\beta\hbar\omega_{q}}-1}\delta\left(\omega+\omega_{q}\right)\right]\\
 &=& e^{\beta\hbar\omega}P_{zt,xy}\left(\omega\right)
 \label{eq. pPi}
\end{eqnarray}

From the definitions in Eq. (\ref{eq. pPi})  it is easy to show that $\Pi$ is not independent function and in fact for all frequencies it can be obtained as follows:
 
\beqr
\Pi_{xy,zt}(\omega)&=&\frac{1}{2\pi} \int d\nu  \frac{S_{xy,zt}(\nu)}{\omega-\nu}
\label{eq. Q S}\\
\Pi_{xy,zt}(-\omega)&=&-\frac{1}{2\pi} \int d\nu  \frac{S_{zt,xy}(\nu)e^{\beta \nu}}{\omega-\nu}
\label{eq. Q S neg w}
\eeqr

Moreover the Fourier transform of the correlator $S_{xy,zt}(\tau)$ defined below Eq. (\ref{rhot})  is 

\begin{eqnarray}\nonumber
S_{xy,zt}\left(\omega\right) & \equiv & \int_{-\infty}^{\infty}\langle X_{xy}\left(t\right)X_{zt}\left(t+\tau\right)\rangle e^{i\omega\tau}d\tau\\
 & = & 2P_{xy,zt}\left(\omega\right)
 \label{eq. S and P}
\end{eqnarray}
  
Generalized dynamical susceptibility is defined below and can be simplified using the $P$ and $\Pi$ functions:

\begin{eqnarray} \nonumber
\chi_{xy,zt}\left(\omega\right) & = & -\frac{i}{\hbar}\int_{-\infty}^{0}\langle\left[X_{xy}\left(t+\tau\right),X_{zt}\left(t\right)\right]\rangle e^{-i\omega\tau}d\tau\\ \nonumber
 & = & \frac{1}{\hbar}\left(\Pi_{xy,zt}\left(\omega\right)+\Pi_{zt,xy}\left(-\omega\right) \right) \\ && -\frac{i}{\hbar}\left[P_{xy,zt}\left(\omega\right)-P_{zt,xy}\left(-\omega\right)\right]
 \label{eq. chi general}
\end{eqnarray}

From above definitions  and after a few lines of simple algebra one can show
\beqr
\chi_{xy,zt}(\omega)=\frac{1}{2\pi\hbar} \int dz \frac{S_{xy,zt}(z)}{\bar{n}(z) (\omega-z)} + \frac{i  S_{xy,zt}(\omega)}{2\hbar \bar{n}(\omega)} 
\label{eq. chi S and int S}
\eeqr

Determining spectral density in terms of susceptibility enables us to simplify the dynamics of density matrix. From eq. (\ref{eq. chi general})  we can split the dynamical susceptibility into symmetric $\chi^+$ and asymmetric $\chi^-$ parts 
\beq
\chi_{xy,zt}(\omega)=\chi_{xy,zt}^+(\omega)+i\tilde{\chi}_{xy,zt}(\omega)
\label{eq. chi to chip and chitilde}
\eeq
where $\chi^{\pm}_{xy,zt}(\omega)\equiv [\chi_{xy,zt}(\omega) \pm \chi_{zt,xy}(-\omega)]/2$ and $\tilde{\chi}_{xy,zt}(\omega) \equiv -i \chi^-_{xy,zt}(\omega)$. The following two identities determine spectral density and $\chi^+$ in terms of $\tilde{\chi}$:
\beqr
S_{xy,zt}(\omega)&=& 2\hbar \bar{n}(\omega)
 \tilde{\chi}_{xy,zt}(\omega)  
 \label{eq. S tilde chi}\\
\chi^+_{xy,zt}(\omega) &=& \frac{1}{\pi} \int dz  \frac{
 \tilde{\chi}_{xy,zt}(z)}{\omega-z} 
 \label{eq. chiplus tilde chi}
\eeqr
and since $\chi_{xy,zt}(\omega)^*=\chi_{yx,tz}(-\omega)$ one can show that 

\beq 
\tilde{\chi}_{xy,zt}(\omega)= \frac{\chi_{xy,zt}(\omega)-\chi^*_{tz,yx}(\omega)}{2i}.
\label{eq. tildechi to chi and chistar}
\eeq

Also in general from eqs. (\ref{eq. pPi}) and (\ref{eq. S tilde chi}) it is simple to prove

\beq
\tix_{xy,zt}(-\omega)=-\tix_{zt,xy}(\omega)
\eeq

Putting the relations in Eq. (\ref{eq. P}) and (\ref{eq. chi general}) one can see that $\tix$ consists of a continuous series of $\delta$-peaks
\beqr \nonumber
\tilde{\chi}_{xy,zt}(\omega)&=&\hbar \pi \sum_q \left[ c^*_{yx,q}c_{zt,q}\delta(\omega- \omega_q)   \right. \\  && \left. \ \  \ \ \qquad -c_{xy,q}c_{tz,q}^*\delta(\omega+\omega_q) \right]
\label{eq. tildechi delta dirac}
\eeqr

Simple algebra shows  $\left[\tilde{\chi}_{xy,zt}(\omega)\right]^*=\tilde{\chi}_{tz,yx}(\omega)$,  indicates the invariance of Eqs. (\ref{eq. S tilde chi}) and (\ref{eq. chiplus tilde chi}) under complex conjugate.  In the example of a two level system using eq. (\ref{eq. tildechi to chi and chistar}) it is simple to show $\tilde{\chi}_{01,10}(\omega)=\imx_{01,10}(\omega)$. Also non-zero spectral density are $S_{01,10}$ and $S_{10,01}$ which are both proportional to $|c_{01}|^2$ and $|c_{10}|^2$ which make the spectral densities to be real. In this case $\imx(\omega)=S(\omega)/2\hbar \bar{n}(\omega)$.


\section{Dynamics in a four-level QHE}
\label{app3}
{To simplify the Bloch equation we consider in the thermally equilibrium heat baths the following relations holds: $\langle \hat{b}^\dagger_{\bm q} \hat{b}_{{\bm q'}}\rangle = \bar{n}_{\bm q} \delta_{{\bm q}{\bm q'}}$ and $\langle \hat{b}_{\bm{q'}} \hat{b}^\dagger_{\bm q} \rangle = \tilde{n}_q) \delta_{{\bm q}{\bm q'}}$ with $\tilde{n}\equiv  (\bar{n}+1)$. In the Weisskopf-Wigner approximation we replace $n_{\bm{q}}\approx n_{{|q|}}$.  In Eq (\ref{eq. eom})  the sum over ${\bm q}$ can be  replaced by an integral through the prescription: $\sum_{\bm q} \to (V_{ph}/\pi^2) \int_0^\infty q^2 dq$ with $V_{ph}$ being the photon volume. In this work we restrict ourselves to steady solutions, therefore we can assume that the density matrix is a slowly varying function of time (Markov approximation) and approximate $\rho(t-\tau)\approx \rho(t)$.  For nearly degenerate levels 1 and 2 the thermal bath can simultaneously exchange energy with both of the levels, therefore we can approximate $n_{{q}_1}\approx n_{{q}_2}\approx n_{{q}}$ with ${q}=(q_1+q_2)/2$. }

 $d\hat{R}/dt=\hat{L}\hat{R}$, defining 

\begin{eqnarray} \nonumber
L\equiv \left(\begin{array}{ccccc}
\chi_{11} & 0 & \tilde{\chi}_{h1}\tilde{n}_{h} & \tilde{\chi}_{c1}\tilde{n}_{c} & -2\tilde{\chi}_{1a,a2}\\
0 & \chi_{22} & \tilde{\chi}_{h2}\tilde{n}_{h} & \tilde{\chi}_{c2}\tilde{n}_{c} & -2\tilde{\chi}_{2a,a1}\\
\tilde{\chi}_{h1}\bar{n}_{h} & \tilde{\chi}_{h2}\bar{n}_{h} & \chi_{hh} & \Omega^{2}n_{l} & 2\tilde{\chi}_{1h,h2}\bar{n}_{h}\\
\tilde{\chi}_{c1}\bar{n}_{c} & \tilde{\chi}_{c2}\bar{n}_{c} & \Omega^{2} \tilde{n}_{l} & \chi_{cc} & 2\tilde{\chi}_{1c,c2}\bar{n}_{c}\\
-\tilde{\chi}_{1a,a2} & -\tilde{\chi}_{2a,a1} & \tilde{\chi}_{1h,h2}\tilde{n}_{h} & \tilde{\chi}_{1c,c2}\tilde{n}_{c} & \chi
\end{array}\right)\\
\end{eqnarray}
with the following definitions: $\chi_{ii}\equiv -\sum_{\alpha} \tilde{\chi}_{i\alpha,\alpha i} \bar{n}_i$,  $\chi_{hh}\equiv -\sum_{i} \tilde{\chi}_{ih,h i} \tilde{n}_h-\tilde{n}_l\Omega^2$, $\chi_{cc}\equiv -\sum_{i} \tilde{\chi}_{ic,c i} \tilde{n}_c- n_l \Omega^2$, $\chi=-\sum_{i,\alpha}\tilde{\chi}_{i\alpha,\alpha i} \bar{n}_\alpha -1/\tau_2$, an the symmetric $\chi_{ij}\equiv \sum_{\alpha}\tilde{\chi}_{i\alpha,\alpha j} \bar{n}_\alpha/2$ for $i\neq j$. 

\section{R/FCS correspondence}
\label{app4}

Consider a QHE is brought to steady state, where energy exchange between the quantum system and the probe environment is measured  over a time $\mathcal{T}$.  A good measure of heat dissipation fluctuations is provided by their full counting statistics (FCS), namely, the full probability distribution for the number of particles or amount of energy exchanged during a given time interval.  The FCS of energy transfers concentrates on the probability $P(E_{tr},\mathcal{T})$ which in the low frequency limit of long $\mathcal{T}$ can generate all statistical cumulants of the energy transfer through the generating function $F(\xi) = \int dE_{tr} P(E_{tr},\mathcal{T}) \exp(i\xi E_{tr}) \approx \exp (-\mathcal{T} f(\xi))$. The parameter $\xi$ is a characteristic parameter and cumulants are given by expansion of $f(\xi)$ in $\xi$ at $\xi=0$. Therefore the cumulant generating function $f(\xi)=-\ln F(\xi)/\mathcal{T}$. There is a Keldysh technique for calculating FCS generating function through an evolution with the following pseudo-density matrix ${\bm R'}$ via the Hamiltonians $\hat{H}^{+,-}$ that are different at forward and backward part of the Keldysh contour\cite{Kamenev,nazarovbook,Kindermann}:
\begin{equation}
\hat{\bm {R'}}(t)  = {\rm T}e^{i\int^t_{-\infty} d\tau \hat{H}^+(\tau)} \hat{\bm R'}(-\infty) {\rm \tilde{T}}e^{-i\int_{-\infty}^t d\tau \hat{H}^-(\tau)}
\label{eq:ExtendedQuantumEvolution}
\end{equation}
Although Eq. (\ref{eq:ExtendedQuantumEvolution}) is very similar to Eq. (\ref{rhot}), the evolution for different Hamiltonians is not unitary. Consequently, $\hat{\bm R'}(t)$ is not a density matrix, in particular, its trace is not $1$.   For two interacting systems A and B with interaction Hamiltonian  $\hat{H}_{AB} = \sum_{xy}\hat{A}_{xy}\hat{B}_{xy}$ we compute the statistics of energy in the subsystem A using the modified $\hat{H}^{\pm}_{AB}(t) = \sum_{xy}\hat{A}_{xy}(t\mp\xi/2) \hat{B}_{xy}(t).$Trace of $\hat{\bm R'}(t)$ defines the statistics of transfers of energy to/from the subsystem $A$. 

Exact analysis in [\onlinecite{ansari2}] shows that there are two types of statistics contributing to the evaluation of the Renyi entropy flow, the generating function of incoherent and coherent energy transfers. The incoherent FCS is \cite{ansari2}:
\begin{equation} 
\label{eq. fi}
\bar{f}^{(T)}\left(\xi\right)  =  - \sum_{x,y,z,t} \int\frac{d\omega}{2\pi}\left(e^{-i\omega\xi}-1\right)S_{xy,zt} ^{\left(\beta\right)}\left(\omega\right)  \mathcal{B}_{xy,zt}(\omega)
\end{equation}
with the definition   
\begin{eqnarray} \nonumber
\label{eq. curlyY}
\mathcal{B}_{xy,zt}\left( \omega \right)  \equiv  {\frac{1}{\mathcal{T}}   \int_{0}^{\mathcal{T}}dt }\int_{-\infty}^{t}dt'  \bigg\{  \left\langle \hat B_{xy}\left(t'\right) \hat B_{zt}\left(t\right)  \right\rangle e^{-i\omega (t-t')}
\bigg. && \\  \nonumber    \bigg. +  \left\langle \hat B_{xy}\left(t\right) \hat B_{zt} \left( t'\right)  \right\rangle e^{i\omega (t-t') }   \bigg\},&&
\end{eqnarray}
with  ${S}_{xy,zt}$ being  standard correlator in the system $A$.

Since a driving force is  externally applied,   another type of energy exchange is possible to take place between the driving force and  the probe environment. One way to consider this energy transfer is to replace $\hat{B}_{xy}$ with the driving energy:  $\hat{B}_{xy} \to \langle \hat{B}_{xy}\rangle$. This leads to coherent FCS similar to Eq. (\ref{eq. fi}). \cite{ansari2} Exact correspondence between the flow of Renyi entropies and the FCS of energy transfers in the weak coupling limits \cite{ansari2} allows for determining  the flow of Renyi entropy  $F_M^{(\beta)}$ for the probe environment as follows:
\beq
F_M^{(T)}/M= f_{incoh}^{(T/M)}(\xi)-f_{coh}^{(T/M)}(\xi)
\label{eq.RFCS}
\eeq
with the characteristic parameter $\xi=i\beta(M-1)$.

\end{document}